# A Deep Learning Model with Radiomics Analysis Integration for Glioblastoma Post-Resection Survival Prediction

Zongsheng Hu[1], Zhenyu Yang[2], Haozhao Zhang[1], Eugene Vaios[2], Kyle Lafata[2,3,4], Fang-Fang Yin[1,2], Chunhao Wang*[2]

[1]Medical Physics Graduate Program, Duke Kunshan University, Kunshan, Jiangsu, China 215316
[2]Department of Radiation Oncology, Duke University, Durham, NC, 27710
[3]Department of Radiology, Duke University, Durham, NC, 27710
[4]Department of Electrical and Computer Engineering, Duke University, Durham, NC, 27710

Short Running Title: Feature Fusion DL for GBM

*Corresponding author:
Chunhao Wang, Ph.D.
Box 3295, Department of Radiation Oncology
Duke University Medical Center
Durham, NC, 27710, United States
E-mail: chunhao.wang@duke.edu

# Abstract

Purpose:

To develop a novel deep learning model that integrates radiomics analysis in a multi-dimensional feature fusion workflow for glioblastoma (GBM) post-resection survival prediction

Methods:

A cohort of 235 GBM patients with complete surgical resection was divided into short-term/long-term survival groups with 1-yr survival time threshold. Each patient received a pre-surgery multi-parametric MRI exam with 4 scans: T1, contrast-enhanced T1 (T1ce), T2, and FLAIR. Three tumor subregions were segmented by neuroradiologists, and the whole dataset was divided into training, validation, and test groups following a 7:1:2 ratio.

The developed model comprises three data source branches: in the $1^{st}$ radiomics branch, 456 radiomics features (RF) were calculated from the three tumor subregions of each patient's MR images; in the $2^{nd}$ deep learning branch, an encoding neural network architecture was trained for survival group prediction using each single MR modality, and high-dimensional parameters from the last two network layers were extracted as deep features (DF). The extracted radiomics features and deep features were processed by a feature selection procedure to reduce the dimension size of each feature space. In the $3^{rd}$ branch, patient-specific clinical features (PSCF), including patient age and three tumor subregions volumes, were collected from the dataset. Finally, data sources from all three branches were fused as an integrated input for a supporting vector machine (SVM) execution for survival group prediction. Different strategies of model design were investigated in comparison studies, including 1) 2D/3D-based image analysis, 2) different radiomics feature space dimension reduction methods, and 3) different data source combinations in SVM input design.

Results:

The model achieved 0.638 prediction accuracy in the test set when using patient-specific clinical features only, which was higher than the results using radiomics features/deep features as sole input of SVM in both 2D and 3D based analysis. The inclusion of radiomics features or deep features with patient-specific clinical features improved accuracy results in 3D analysis. The most accurate models in 2D/3D analysis reached the highest accuracy of 0.745 with different combinations of dissimilarity-selected radiomics features, deep features, and patient-specific clinical features, and the corresponding ROC area-under-curve (AUC) results were 0.69 (2D) and 0.71 (3D), respectively.

Conclusion:

The integration of radiomics features, deep features, and patient-specific clinical features in the designed model improves post-surgery GBM survival prediction.

**Key Words**: GBM, radiomics, deep learning, SVM, MRI

# Introduction

Arising from glial cells, glioblastoma (GBM) is the most common primary malignant brain tumor, accounting for 16% of all central nervous system neoplasms[1]. The average age-adjusted incidence rate of GBM is 4.43 per 100,000 population[2]. In the meantime, GBM is one of the most fatal cancers with short overall survival (OS): over 70% of GBM patients will suffer disease progression within one year after diagnosis[3], with a five-year survival rate of less than 5% [2].

For individualized patient management in clinical practice, accurate and reliable predictions of GBM patient survival are highly demanded to enable early treatment interventions and alternative strategy exploration; however, such survival prediction remains challenging. Studies showed that GBM patients with the same tumor histopathology may have significantly different survivals, which renders the histopathomic approach qustionable[4]. Radiography analysis, particularly MRI-based image analysis, aims at decoding radiographic phenotype of tumor and normal brain from high soft tissue contrast at millimeter level, and thus becomes an active field for GBM OS prediction[5]. In particular, quantitative MR image analysis via computational approach, represented by radiomics techniques, extracts high-level descriptive statistics beyond typical image rendering dimensions. These statistics were hypothesized to benefit GBM survival prediction in combination with advanced statistical modelling[6,7]. For instance, a representative study by Kickingereder et al extracts radiomics features from multi-parametric anatomical MRI exams for supervised principal analysis, and this approach outperformed classic risk model in low risk/high risk patient group stratification[8]. Other radiomics studies have studied quantitative MRI protocol use, genetic profile integration[9], and novel statistical approaches[10] for enhanced GBM survival prediction. Recently, deep learning (DL) approaches have been investigated for MR-based GBM OS prediction. Unlike classic radiomics analysis that requires handcrafted feature extraction with experimental knowledge, deep learning models calculate a very large number of features simultaneously, while the risk modelling via deep neural network is supported by massive computations without explicit expressions[11]. To date, pilot works have reported GBM survival group stratification and regression based on deep learning designs[12-14].

The integration of radiomics analysis and deep learning implementation for quantitative image analysis is conceptually appealing: on one hand, quantitative information extracted by deep neural network can potentially work jointly with radiomics features in a complementary way to improve risk modelling performance; on the other hand, potential interactions between radiomics features as handcrafted descriptors with known expressions and complex deep features may provide a radiomics aspect of deep learning model explanation to address well-known 'black-box' issue[15]. However, no such studies have been studied for MR-based GBM survival prediction. One of the major technical obstacles is data dimensionality design: while convolution operations in deep neural networks are suitable for spatial image manipulation, direct interactions between low dimensional radiomic feature vectors and neural network will be problematic due to network dimension overflow and potential overfitting. In this work, we establish a novel multi-dimensional deep learning design that enables radiomics analysis integration. It features two innovative solutions: 1) classic radiomics features and deep features, i.e., latent variables of feature extraction layers in a deep neural network, are extracted simultaneously from a multi-parametric MRI protocol; and 2) radiomics features and relevant non-image clinical features in lower dimensions are connected to the deep features in an abstract feature space, and such connected feature is utilized by Support Vector Machine (SVM)[16] as a supervised machine learning model for GBM survival risk group prediction. We demonstrate our development using a publicly available GBM image dataset, and we conducted multiple comparison studies to study its potential merits.

# Materials and Methods

*Image Data*

In this study, we studied a total of 235 GBM patients from the BraTS 2020 database[17,18]. Each patient has a pathologically confirmed diagnosis and a pre-operative multi-parametric MRI exam that included four series: T1 weighted, T2 weighted, contrast-enhanced T1 weighted (T1ce), and water-suppressed FLAIR (Fluid Attenuation Inversion Recovery). All four MR series have been co-registered with skull-skipping processing, and image resolutions have been unified to isotropic 1 mm$^3$. In addition to survival in days following complete resection surgery, each patient has ground-truth tumor segmentation results from experienced neuro-radiologist contouring. These segmentation results include three key regions: contrast-enhanced tumor region, peritumoral edema, and the necrotic/non-enhancing tumor core. In this work, these patients were divided into two groups: 116 patients in the short-term survival group (<1yr survival) and 119 patients in the long-term survival group (>1yr survival) after complete resection surgery. Such group assignment follows most clinical relevance for a balanced data group assignment.

*Model Design*

Figure 1 summarizes the overall design of our radiomics-integrated deep learning workflow for GBM survival group prediction. As illustrated, multi-dimensional data processing comprises three branches. In the 1$^{st}$ branch, radiomics features are calculated using each MR modality from selected regions-of-interest (ROIs), and dimension reduction methods are adopted to select condensed radiomics features from the whole feature space. In the 2$^{nd}$ branch, an encoding neural network architecture is trained using each single MR modality for short-term/long-term survival prediction. Higher level features from last two dense layers are extracted as deep features, followed by dimension reduction processing. In the 3$^{rd}$ branch, the patient-specific clinical features (PSCF) were collected from the dataset. Due to the public dataset content, we collected patient age at the time of surgery and calculated 3 tumor region volumes (i.e., contrast-enhanced region, non-enhancing region, and peritumoral edema) based on ground-truth segmentation results. Finally, data from all three branches are fused as an integrated input for SVM modelling to predict the survival group. In this work, we investigated different strategies of the designed model implementation, including 1) 2D/3D approaches of image analysis; 2) different dimension reduction strategies; and 3) different combinations of SVM input design from the three branches. The whole patient dataset was divided into training, validation, and test set following a ratio of 7:1:2.

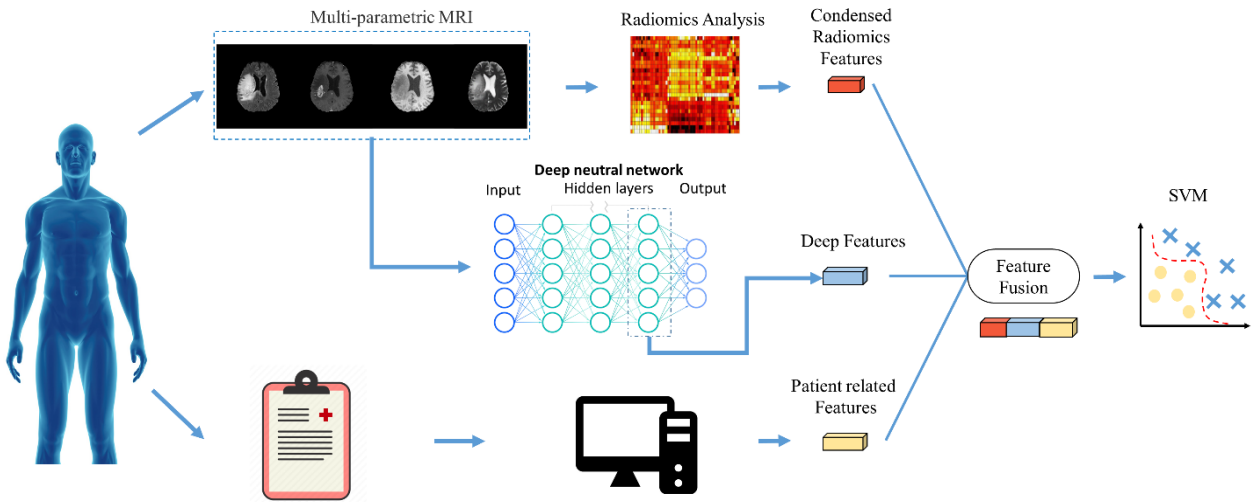

*Figure 1. A flowchart illustration of the radiomics-integrated deep learning model. A total of 3 data branches are included, top: radiomics features; middle: deep features; bottom: patient-specific clinical features were extracted and compressed as the fussed input of SVM*

A. Radiomics Analysis

In this work, we extracted 22 textural features from Gray Level Co-Occurrence Matrix (GLCOM)[6] and 16 textural features from Gray Level Run Length Matrix (GLRLM)[7] using our in-house radiomics software package[19,20], and features from three segmentation regions, i.e., contrast-enhanced tumor region, peritumoral edema, and the necrotic tumor core, were collected. In 2D based analysis, the axial slice with the biggest cross-sectional area of the contrast-enhanced tumor region was selected, and radiomics features were extracted from the three segmented regions in the selected slice. In 3D based analysis, radiomics features were extracted from the 3D segmentation of the regions with isotropic resolution. Therefore, 456 radiomics features ((22+16) features x 4 MR modalities x 3 regions) were derived from each patient in 2D/3D analysis. It is a known issue that radiomics features may demonstrate high correlations with each other[21]. As such, proper dimension reduction methods are in demand to reduce the influence of repetitive information and overfitting possibilities. In this work, we studied two approaches in feature selection: principal component analysis (PCA) and dissimilarity analysis. In the PCA approach, 14 main components were obtained for maintaining >90% explain ratio of original full radiomics feature space. In dissimilarity analysis, radiomics features that maximized differences between long-term and short-term survival groups were selected based on the defined dissimilarity below:

$$Dissimilarity = \frac{|mean(L) - mean(S)|}{|mean(L)| + |mean(S)|}$$

where mean(S) and mean(L) refer to the mean values of the feature over short-term and long-term survival patient groups, respectively. 7 radiomics features were selected with dissimilarity over the threshold of 0.15.

B. Deep Feature Extraction

In this work, we adopted different deep neural network designs in the 2$^{nd}$ branch of Figure 1 to accommodate 2D and 3D based image analysis. Figure 2A summarizes the deep neural network for 2D image inputs. Based on a pre-trained VGG-16 architecture[22], the network consists of two parts: the 1$^{st}$ part is a convolutional base with 5 convolutional blocks. Each convolutional block is stacked by 2 or 3 convolutional layers and a max-pooling layer. In each convolutional layer, the filter size is 3×3 with padding and stride of 1. Max-pooling is performed over a 2x2-pixel window with a stride of 2. The 2$^{nd}$ part is the dense part, which is the stack of 5 dense layers with the size of 1024, 1024, 512, 256, and 3, respectively. To avoid the occurrence of overfitting, a dropout layer was added between the first two Dense layers with a dropout possibility of 0.5, and soft-max activation was used in the output layer. The input of the neural network is a three-channel image with a 160×208×3 shape size following the BraTS dataset specification, while the output is categorical binary label vectors, i.e., [1,0] and [0,1], which correspond to short-term and long-term survival groups, respectively. To deal with relatively small data size in this work, the convolutional base loaded the weights that were pre-trained on ImageNet[23] as a transfer learning scheme[24].

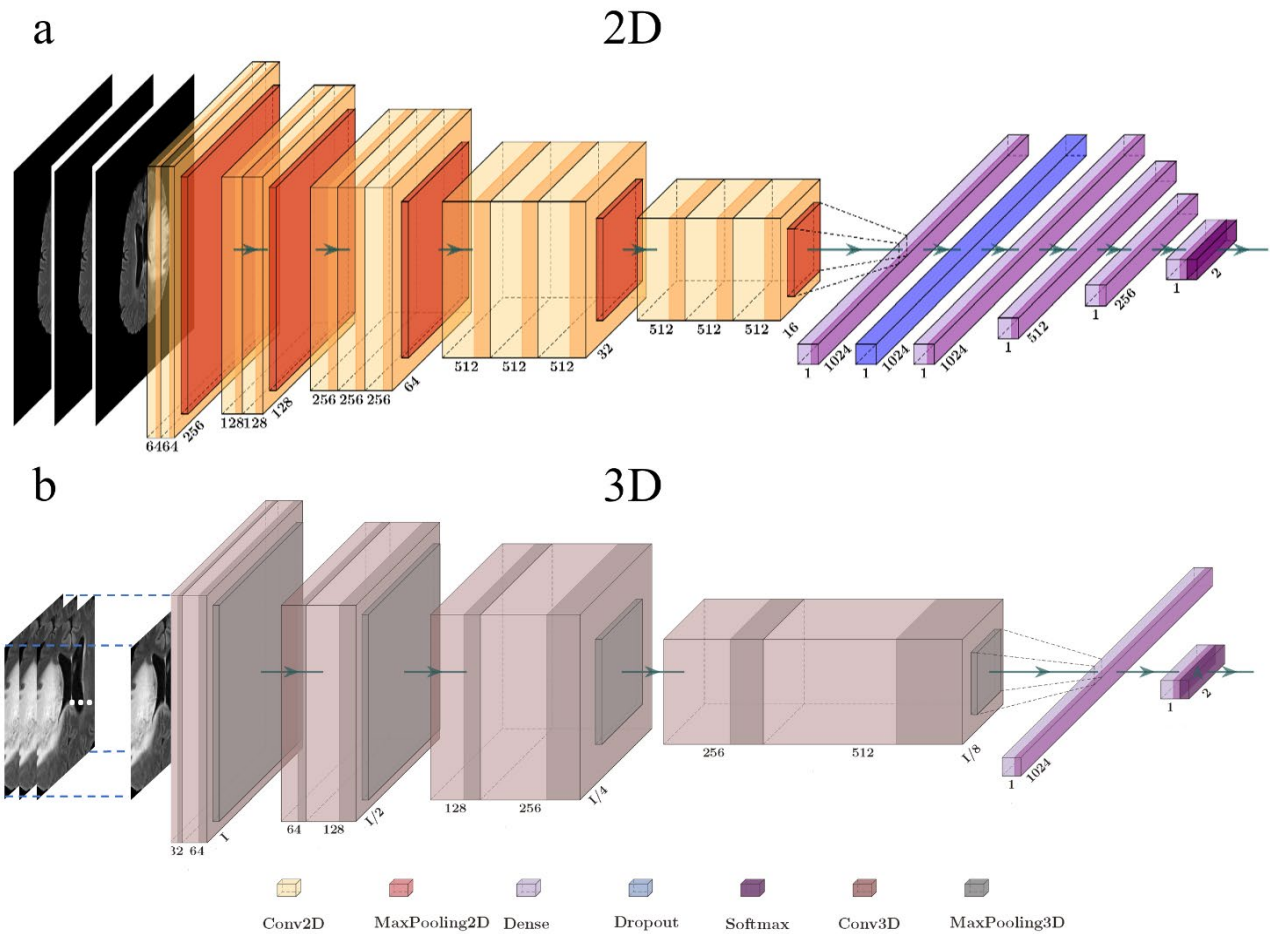

Figure 2B summarizes the deep neural network for 3D image inputs. Similarly, the proposed 3D CNN architecture consists of a convolutional base and dense part. The convolutional base is constructed based on the encoding part of U-Net architecture[25]. The convolutional base is loaded with the weights that are pre-trained with the segmentation task on a medical imaging dataset[26,27]. The convolutional base consists of 4 convolutional blocks. Each convolutional block is stacked by 2 or 3 3D convolutional layers and a 3D max-pooling layer. In each convolutional layer, the filter size is 3 x 3 x 3 with padding and stride of 1. Max-pooling is performed over a 2 x 2 x 2-pixel window with a stride of 2. The dense part is the stack of 2 dense layers with a size of 1024 and 2. Soft-max activation was used in the output layer.

In both 2D and 3D scenarios, ten network versions were trained with randomized training and validation assignment following a 7:1 ratio. All images were normalized to [0, 1] range before training. During training, Adam optimizer was selected, and the loss function was categorical cross-entropy during training. Basic data augmentation strategies of image translation and rotation were included. Deep features were extracted from the last two dense layers of a network version with the highest (of the 10 versions) validation accuracy. Similarly, the PCA method was implemented as dimension reduction processing: in the 2D scenario, 2/3/2/4 main components from the network trained with FLAIR/T1/T1ce/T2 images were extracted with explain ratio threshold 0.90, respectively; similarly, in the 3D scenario, 3/3/3/5 main components were extracted as FLAIR/T1/T1ce/T2 parameters.

C. SVM Design

The extracted features in Figure 1 were concatenated as a fused feature, an SVM model was trained to predict short-term/long-term survival group results. Prior to SVM training, all integrated inputs were normalized to [-1, 1] range. During SVM training, a linear kernel was selected to enable studies of input components' contribution to final prediction results. The C value, which represents misclassifying tolerance during training, was set to $40^{28}$.

*Comparison Studies*

We evaluated the performance of the proposed deep learning design in a two-fold comparison study design. First, we compare GBM survival group prediction accuracies using sole data source, i.e., deep learning network results, radiomics modelling results, and PSCF modelling results. A same SVM design was used for radiomics modelling and PSCF modelling. As the next step, we compared prediction accuracies of our proposed deep learning model with different radiomics dimension reduction strategies and different integrated feature designs of SVM input in Figure 1. Both 2D and 3D scenarios were studies.

# Result

Table 1 summarizes the survival prediction results from vanilla deep learning results, i.e., classification results from deep neural networks' output in Figure 2. In general, when using single MR modality only, deep neural network results in the test set are not promising and the average accuracy result is around 0.5. The best single-input result using 3D FLAIR volumes achieved an accuracy of 0.570 in the test set, but the accuracy in the training set was not satisfying (= 0.54). As a reference, when multiple MR modalities were used as deep neural network input, the combination of FLAIR, T1ce, and T2 were found with the best and yet unsatisfying performances in both 2D and 3D execution. In sum, the current results suggest that GBM survival group prediction based on classic MR-based deep learning implementation is limited; this observation is consistent with the challenging GBM survival group prediction task in clinic.

| Input | 2D | | 3D | |
|---|---|---|---|---|
| | Train Accuracy | Test Accuracy | Train Accuracy | Test Accuracy |
| Flair | 0.859 | 0.509 | 0.540 | 0.570 |
| T1 | 0.772 | 0.502 | 0.565 | 0.483 |
| T1ce | 0.897 | 0.545 | 0.513 | 0.568 |
| T2 | 0.913 | 0.551 | 0.604 | 0.436 |
| FLAIR+T1ce+T2 | 0.855 | 0.530 | 0.521 | 0.542 |

*Table 1. Survival group prediction results of deep neural network output as in Figure 2.*

The results of SVM execution in Figure 1 using single branch, i.e., radiomics feature ($RF$), deep feature ($DF$), or PSCF (age + tumor region volumes), are summarized in Table 2. When deep features were used, the best accuracy results in the test set were 0.660 in 2D scenario ($DF_{FLAIR}+DF_{T1ce}+DF_{T2}$) and 0.596 in 3D scenario ($DF_{FLAIR}+DF_{T1}+DF_{T2}$); these results are improved from Table 1 results. Note that 2D scenario result is better than 3D scenario result. When radiomics features were used, features selected by dissimilarity analysis ($RF_{DA}$) outperformed the features selected by PCA ($RF_{PCA}$) in both

2D and 3D scenarios. When PSCF was used, the prediction accuracy in test accuracy was 0.638, which is better than results from radiomics feature utilization.

| Input | 2D | | 3D | |
|---|---|---|---|---|
| | Train Accuracy | Test Accuracy | Train Accuracy | Test Accuracy |
| $DF_{FLAIR} + DF_{T1(ce)} + DF_{T2}$ | 0.984 | 0.660 | 0.718 | 0.596 |
| $RF_{PCA}$ | 0.622 | 0.511 | 0.622 | 0.511 |
| $RF_{DA}$ | 0.601 | 0.574 | 0.590 | 0.596 |
| PSCF | 0.697 | 0.638 | 0.697 | 0.638 |

*Table 2. Survival group prediction results of SVM implementation with single data source branch in Figure 1. DF: deep feature; RF: radiomics feature; TS: tumor size*

Table 3 summarizes the key results of the proposed deep learning model in Figure 1 with multiple data branch sources. When radiomics features were utilized together with PSCF, the test group accuracy result was improved from 0.638 (Table 2) to 0.681 in 3D scenario. Similarly, when deep features were utilized together with PSCF, the accuracy was further improved to 0.723. When deep features were included together with PSCF and radiomics features, the achieved test group accuracy was further improved to 0.745 in both 2D scenario ($PSCF + RF_{DA} + DF_{T1ce}$) and 3D scenario ($PSCF + RF_{DA} + DF_{FLAIR} + DF_{T1ce} + DF_{T2}$). The corresponding ROC curves of the identified two models are illustrated in Figure 3. Both curves exhibited similar shapes, and the area-under-curves (AUC) values were 0.69 for the 2D model curve (blue) and 0.71 for the 3D model curve (red). In a nutshell, both 2D and 3D implementation of the proposed multi-dimensional deep learning model achieved the same level of accuracy in GBM survival group prediction.

| Input | 2D | | 3D | |
|---|---|---|---|---|
| | Train Accuracy | Test Accuracy | Train Accuracy | Test Accuracy |
| $PSCF + RF_{DA}$ | 0.622 | 0.511 | 0.718 | 0.681 |
| $PSCF + DF_{FLAIR}$ | 0.697 | 0.638 | 0.718 | 0.723 |
| $PSCF + RF_{DA} + DF_{T1ce}$ | **0.941** | **0.745** | 0.707 | 0.702 |
| $PSCF + RF_{DA} + DF_{T2}$ | 0.957 | 0.638 | 0.755 | 0.617 |
| $PSCF + RF_{DA} + DF_{FLAIR} + DF_{T1ce} + DF_{T2}$ | 0.984 | 0.574 | **0.777** | **0.745** |

Table 3. Survival group prediction results of SVM implementation with multiple data source branches in Figure 1. DF: deep feature; RF: radiomics feature; TS: tumor size. The best results in 2D and 3D scenarios are highlighted.

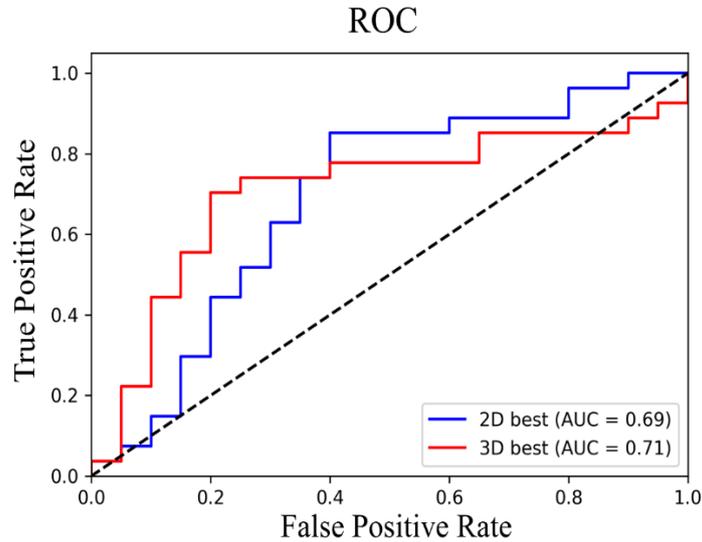

Figure 3. The ROC curves from the identified multi-dimensional deep learning model with the best accuracy results in 2D (blue) and 3D (red) scenarios.

To evaluate the relative contribution of different features to the final prediction results in SVM execution, we calculated the SVM coefficient of each feature from the aforementioned best models in 2D and 3D scenarios. These results are shown in Figure 4. In general, all three categories' features demonstrated certain contributions to the final prediction results. Although some contribution differences from three categories are observed, none of the categories made dominant contributions in both 2D and 3D scenarios. The feature with the highest coefficient was found within deep feature group in 2D scenario and within radiomics feature group in 3D scenario, respectively. These results suggest that the proposed model utilized the information from different data source branches in a complementary way without unbalanced focus.

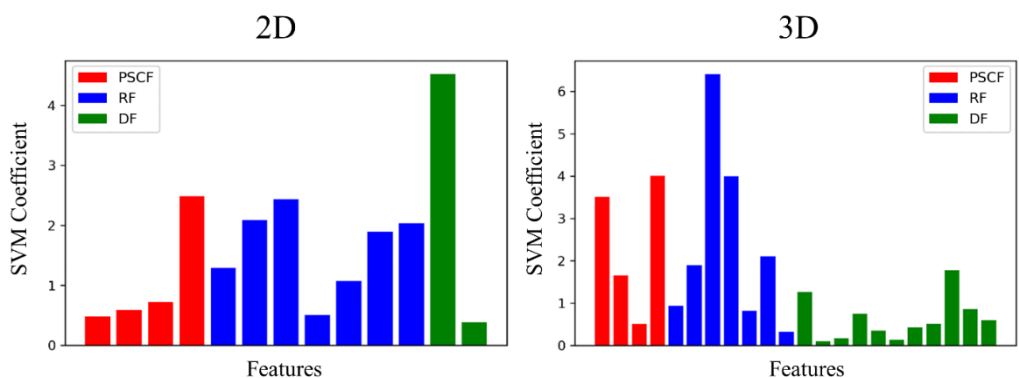

*Figure 4. SVM coefficients of features in 2D (left) and 3D (right) models with the best prediction accuracy results. PSCF: patient-specific clinical feature (age and tumor size); RF: radiomics feature; DF: deep feature*

# Discussion

In this work, we developed a novel deep learning model that enables radiomics analysis integration for treatment outcome prediction. To our best knowledge, this work is the first of its kind in multi-dimensional feature fusion for GBM survival group prediction. As handcrafted computational features with known expression, radiomics features reflect quantitative evaluations of high-order statistics that describe image intensity distribution patterns. On the other hand, deep features embedded in a deep neural network learnt image-specific patterns in high-dimensional feature space automatically without analytical expression. The two feature types were hypothesized to possess complementary information due to different feature extraction processes, and thus the combination of the two types in a single machine learning model can potentially improve MR-based GBM survival group prediction with their different extraction. In current clinical practice, basic patient information, including patient age at surgery, tumor location, and tumor size, are reasonable indicators of GBM post-surgery survival time[29-33]. Our results in Table 2 reported that when using age + tumor region volumes only, the survival group prediction accuracy reached a decent level of 0.638. Results showed that the joint use of radiomics features, deep features, and patient-specific clinical features further improved prediction accuracy to 0.745, suggesting that the great potential of the current study design in clinical practice. It is worth mentioning that the best models in 2D and 3D scenarios achieved the same accuracy of 0.745 in Table 3 and very close ROC AUC results in Figure 3 (2D: 0.69; 3D: 0.71). This result suggests that 2D-based model execution, though with different feature selection results, could be equivalent to 3D-based model execution in terms of final results. Intuitively, 3D-based model design is believed to be superior to 2D-based model design with additional volumetric information. With additional high-throughput computation capability, deep neural network may extract sufficient radiography information based on the central 2D axial slice, particularly for GBM patients with large tumor sizes and discernable radiograph patterns. Thus, the 2D execution of the proposed deep learning model could reach the same performance as in 3D execution. For potential clinical application, 2D execution of the current model design might be favored with its lighter model design and less computational load, which are more important when working with larger patient datasets.

Currently, we studied the GBM survival prediction as a bi-class classification problem with 1-yr as short-term/long-term group threshold. This threshold was set based on two reasons: first, it generally agrees with the clinical observation that many patients demonstrated progression after 1-yr survival[5]; and second, it ensures approximately balanced group sizes for radiomics analysis and deep learning implementation. It would be appealing to study survival group prediction as a regression problem to indicate results in terms of days/months. Due to the limited sample size, however, the regression problem design could not converge to extract discernable patterns other than population average results. In future works with enlarged dataset either from a single institution or from multi-institution collaborations, we plan to extend the current feature fusion architecture in regression problem settings.

In radiomics analysis as presented in Figure 1, two different dimension reduction approaches were investigated to condense radiomic information. Results in Table 2 showed that results by dissimilarity analysis (DA) outperformed principal component analysis (PCA) in both 2D and 3D scenarios. The adopted dissimilarity metric intends to select a group of features with less common information to emphasize their complementary roles in modelling. Meanwhile, the extracted principal components in PCA no longer preserve the original form of radiomics features, which may degenerate the embedded analytical information of radiomics features during SVM execution. Thus, dissimilarity analysis may be a better candidate of dimension reduction technique for radiomics features. During deep feature dimension reduction, dissimilarity analysis was not studied because deep features did not demonstrate meaningful numerical values of dissimilarity, and thus feature selection was not successful; this can be attributed to the fact that deep features are high dimensional statistics without analytical expression, and thus the value-based dissimilarity as $1^{st}$ order statistics is too simple to summarize deep feature patterns. Additionally, deep features are supposed to capture latent differences between the two patient groups during the training, and the dissimilarity measurement may redundantly emphasize the data inhomogeneity. Nevertheless, it might be too immature to rule out the data sample size issue here. When an enlarged dataset becomes available, deep feature dimension reduction will be scrutinized as additional comparison studies.

Besides GBM survival group prediction, the current deep learning model design can be generalized to other oncology applications of outcome prediction based on the technical premise of radiomics analysis and deep learning applicability. In addition, more data source branches, such as genetic profile (i.e., radiogenomic integration[34]), can be added to the current design in Figure 1 as a 'Multi-omics' solution. We anticipate such works may happen soon when relevant data can be available as in potential multi-institutional collaboration.

# Conclusion

In this work, we successfully developed a novel deep learning model that integrates radiomics analysis in a multi-dimensional feature fusion workflow. In a clinical problem of GBM post-resection short-term/long-term survival prediction, results showed that the joint use of radiomics features, deep features, and patient-specific clinical features achieved a prediction accuracy of 0.745. The developed model possesses great potential in future works with additional data source handling before emerging clinical applications.